**ORIGINAL ARTICLE**

# The First Stellar Parallaxes Revisited

Mark J. Reid[1] | Karl M. Menten[2]

[1]Center for Astrophysics | Harvard & Smithsonian, Cambridge, MA, USA

[2]Max-Planck-Institut für Radioastronomie, Bonn, Germany

**Correspondence**
Mark J. Reid, Center for Astrophysics | Harvard & Smithsonian, 60 Garden Street, Cambridge, MA 02138, USA Email: mreid@cfa.harvard.edu

We have re-analyzed the data used by Bessel, von Struve, and Henderson in the 1830s to measure the first parallax distances to stars. We can generally reproduce their results, although we find that von Struve and Henderson have underestimated some of their measurement errors, leading to optimistic parallax uncertainties. We find that temperature corrections for Bessel's measured positions are larger than anticipated, explaining some systematics apparent in his data. It has long been a mystery as to why von Struve first announced a parallax for Vega of $0.''125$, only later with more data to revise it to double that value. We resolve this mystery by finding that von Struve's early result used two dimensions of position data, which independently give significantly different parallaxes, but when combined only fortuitously give the correct result. With later data, von Struve excluded the "problematic" dimension, leading to the larger parallax value. Allowing for likely temperature corrections and using his data from both dimensions, reduces von Struve's parallax for Vega to a value consistent with the correct value.

## 1 | INTRODUCTION

In 1838, Friedrich Wilhelm Bessel published two papers reporting what has generally been considered to be the first highly significant measurement of the distance to a star. One paper published in the *Monthly Notices of the Royal Astronomical Society* (Bessel, 1838c) was an edited version, translated from German to English by John Herschel, of a more detailed paper in *Astronomische Nachrichten* (Bessel, 1838a) titled "Determination of the distance to the 61st star in Cygnus." These papers reported what has generally been considered to be the first highly significant measurement of the distance to a star. Bessel used the Earth's orbit as a surveyor's baseline to measure the apparent angular shift of a nearby star as the Earth moves around the Sun (trigonometric parallax). This achievement culminated centuries of efforts to determine what was effectively a scale size of the Universe for 19th century astronomy, as well directly validating that the Earth went around the Sun. Fernie (1975) and Hirshfeld (2001) place this work in a historical context, detailing the lives and efforts that lead this momentous result.

In the same time period, two other astronomers also collected astrometric data that would yield trigonometric parallax distances to stars. In Cape Town, South Africa, Thomas Henderson observed $\alpha$ Centauri with a mural circle between 1832 and 1833, as part of a program to measure the positions of large numbers of stars. Years after his return to England, he analyzed this data and claimed to have a parallax measurement for what is now known to be the nearest stellar system (Henderson, 1840). In 1835, Friedrich Georg Wilhelm von Struve started observations designed specifically to measure a trigonometric parallax for $\alpha$ Lyrae (Vega), and, after collecting a year's data, he communicated a tentative result in a chapter of his monograph "Stellarum duplicium et multiplicium mensurae micrometricae" (von Struve, 1837).

Interestingly, both von Struve and Bessel used telescopes designed by Joseph Fraunhofer specifically for parallax measurement. These were the last telescopes built by Fraunhofer and in the hands of these two astronomers produced outstanding results, providing a prime example for the validity of the paradigm that major discoveries follow major technological innovations in observational instruments as recognized by Harwit (1981).

Motivated by his friend von Struve's tentative result, Bessel, who had worked on the parallax of 61 Cygni as early as 1812, but had to await a better telescope (Ashbrook, 1954), started an intensive observing program to measure the parallax of this star and beat von Struve to the prize. Given this complicated and interwoven sequence, attribution of priority for this truly transformational result has been a subject of discussion to this date.

We have been developing radio astronomical techniques in order to directly map the spiral structure of the Milky Way. This endeavor involves measuring trigonometric parallaxes for large numbers of massive stars that have recently formed in giant molecular clouds and trace spiral structure throughout the Galaxy. This prompted our interest in the history of the first parallax measurements. Fortunately, much of the raw data has been published, and this paper presents a re-analysis, made possible by modern computers, in order to better assess their significance.

## 2 | FRIEDRICH WILHELM BESSEL

Bessel chose 61 Cygni for a parallax measurement because it could easily be observed over a full year at the Königsberg Observatory in Prussia and, importantly, it was known to have an extremely large motion across the sky (proper motion), suggesting it was nearby. He used a heliometer telescope built specially for high precision astrometry by Joseph Fraunhofer, the world's premier telescope maker. This was the last telescope designed by Fraunhofer, and it was delivered to Bessel after Fraunhofer's death. A heliometer has the objective lens cut in half and the two pieces can be slid along side each other to superpose the images of two stars. Their angular separation can then be determined by a precision micrometer screw, which tells how far one lens was slid with respect to the other.

61 Cygni is a binary system, with the two stars having approximately a 700-year orbital period. Bessel gives their separation in 1838 as $16''$ along a position angle of 95° East of North. He measured the angular distance from the center of the binary to each of two dimmer reference stars: star $a$ separated by $462''$ at a position angle of 201° and star $b$ separated by $706''$ at a position angle of 109°. Bessel took upwards of 100 measurements of the position of 61 Cygni relative to each reference star over one year and performed a least-squares fit to estimate a parallax and a residual proper motion[1]. His parallax, obtained by combining the results for both reference stars, was $0.''314 \pm 0.''020$, where the uncertainty assumes that individual measurements have errors near $\pm 0.''14$ and that the dim reference stars are at great distances so as to not have significant parallaxes of their own. The true parallax of 61 Cyg, as recently measured by *Gaia*, a European Space Agency parallax mission, is $0.''28615 \pm 0.''00006$ (Gaia Collaboration, 2018).

Bessel's individual and combined parallax results are listed in Table 1. Our re-analyses, assuming, as did Bessel, a constant measurement uncertainty, yields nearly identical parallax estimates (see column 4 of Table 1). Plots of data and fitted parallax and proper motion are shown in Fig. 1. Clearly Bessel was able to perform the extremely laborious calculations for a least-squares fit of $\approx 100$ measurements manually without error.

The difference in the parallax estimates of 61 Cyg measured against reference star $a$ and star $b$ is $0.''109 \pm 0.''040$. This difference appears statistically significant ($2.7\sigma$) and, indeed, Bessel noted this tension (Bessel, 1838c):

> "The observations seem also to indicate that the difference of the parallaxes of 61 and $b$ is smaller than that of 61 and $a$; which must be the case, indeed, if $b$ itself have a sensible parallax greater than $a$."

Bessel was pointing out that each result is a *relative* parallax, which measures the difference in true parallax of 61 Cyg and that of a reference star, and he was suggesting that the (unknown) parallax of reference star $b$ could perhaps be significant. For example, were star $b$'s parallax near $0.''1$, that would remove the tension between the two measures. However, *Gaia DR2* parallaxes are $0.''0024$ for star $a$ (BD+37 4173) and $0.''0020$ for star $b$ (BD+37 4179). So both reference stars are effectively at "great distances" and did not significantly bias either parallax measurement.

In Fig. 1, one can see that while the model (solid line) for star $a$ fits the data quite well, this is not the case for star $b$. The residual differences between the measurements and a best fitting parallax (and proper motion) are shown in Fig. 2 and for star $b$ reveal systematic problems. Specifically, one can see that nearly all residuals in the autumn/winter period between 1837.8 and 1838.3 are negative, while those in the spring/summer period between 1838.3 and 1838.6 are mostly positive. This suggests a temperature related problem. Indeed, John Herschel, who translated Bessel's 1838 announcement paper for the *Monthly Notices of the Royal Astronomical Society*, was concerned about possible temperature issues (Bessel, 1838b).

---

[1] Bessel removed the known large motion of the binary system from his angular offset measurements.

**TABLE 1** Parallax of 61 Cygni.

| Star | Data Used | Parallax Bessel | Parallax Re-analysis | Parallax T-adjusted |
| --- | --- | --- | --- | --- |
| 61 Cyg | Reference star $a$ | $0\overset{''}{.}369 \pm 0\overset{''}{.}028$ | $0\overset{''}{.}368 \pm 0\overset{''}{.}028$ | $0\overset{''}{.}270$ |
| 61 Cyg | Reference star $b$ | $0\overset{''}{.}260 \pm 0\overset{''}{.}028$ | $0\overset{''}{.}259 \pm 0\overset{''}{.}028$ | $0\overset{''}{.}297$ |
| 61 Cyg | Combined | $0\overset{''}{.}314 \pm 0\overset{''}{.}020$ | $0\overset{''}{.}313 \pm 0\overset{''}{.}020$ | $0\overset{''}{.}286$ |

Comparison of parallax measurements of the center of the 61 Cyg binary by Friedrich Wilhelm Bessel and two modern re-analyses of his data. The "Re-analysis" values assume a constant measurement uncertainty, with the parallax uncertainty scaled to give a reduced chi-squared per degree of freedom on unity. The "T-adjusted" values are from Bessel's extended data through 1840, using his "temperature correction" formulae for parallax. We have set his temperature parameter $k = 1.17$ to give a combined result that matches the true parallax of 61 Cyg of $0\overset{''}{.}286$.

Fortunately, Bessel listed the temperature corrections, $\Delta_T$, that he applied to his data in his followup 1838 paper in the *Astronomische Nachrichten*. We investigated the effects of these corrections on estimates of parallax by adding to the model a scaled version of the corrections. The scaling was controlled by an adjustable parameter and optimized in the least-squares fitting. Interestingly, we find that the fit can be greatly improved by adding $1.03(\pm 0.18) \times \Delta_T$ to the data for star $b$. The revised parallax estimate for star $b$ is $0\overset{''}{.}213 \pm 0\overset{''}{.}026$. This fit has a $\chi^2 = 74.8$ for 94 degrees of freedom, considerably improved compared to a $\chi^2 = 101.6$ for 95 degrees of freedom for no temperature adjustments. If, instead, we use the modern value for parallax of $0\overset{''}{.}286$ as a strong prior, and solve for the scaling parameter, we obtain a value of 0.88, which is consistent with our fitted value of 1.03 within its formal uncertainty.

Clearly, temperature corrections could have had a significant effect on Bessel's parallax estimates. Applying the same fitting procedure to the data for star $a$, yields a highly uncertain scaling parameter of $-0.50 \pm 0.41$ (note the opposite sign compared to star $b$). If, instead, we adopt the better determined temperature scaling parameter (1.03) from star $b$, we find a parallax of $0\overset{''}{.}291 \pm 0\overset{''}{.}030$ for the star $a$ data. The excellent agreement with the modern parallax value is, perhaps, fortuitous, but this correction certainly improves the value of $0\overset{''}{.}369 \pm 0\overset{''}{.}028$ that Bessel published for this star.

Bessel continued his observations beyond 1838 and published extended results in Bessel (1840a) and Bessel (1840b). In these papers, he explicitly confronts the issue of temperature corrections, giving formulae for parallaxes that include a temperature parameter ($k$):

$$61 \text{ Cyg} - \text{star } a \ldots \quad \pi = 0\overset{''}{.}3584 - 0\overset{''}{.}0758k, \quad (1)$$

$$61 \text{ Cyg} - \text{star } b \ldots \quad \pi = 0\overset{''}{.}3289 - 0\overset{''}{.}0276k, \quad (2)$$

and a combined result

$$61 \text{ Cyg} - \text{stars } a \& b \ldots \quad \pi = 0\overset{''}{.}3483 - 0\overset{''}{.}0533k. \quad (3)$$

Bessel states, $k$ is "a small indeterminate correction depending on the effects of temperature on the micrometer-screw." He points out that "On deducing the value of $k$ from the observations, those of the first star give, therefore, $k = -0.489$; and those of the second, $k = 0.054$." Above, we noted that when we solved for a temperature scaling parameter, for star $a$ we obtained $-0.50$, essentially the same value as Bessel found. However, for star $b$ Bessel's scaling parameter of 0.054 is markedly different from our value of 1.03.

Knowing the true parallax of 61 Cyg of $0\overset{''}{.}286$, we can use the Bessel (1840a) parallax formulae, Eqs. (1), (2) and (3), to assess the consistency of the inferred parallaxes for the two stars, based on Bessel's extended data set. The combined formula given by Eq. (3), sets $k = 1.17$. Adopting this temperature correction, yields individual parallaxes for star $a$ and star $b$ of $0\overset{''}{.}270$ and $0\overset{''}{.}297$. Assuming uncertainties of $\pm 0\overset{''}{.}028$ (see Table 1), these are statistically consistent. Thus, in hindsight, it appears that temperature corrections were larger than Bessel anticipated, and this can explain his small overestimate of the parallax of 61 Cygni.

## 3 | FRIEDRICH GEORG WILHELM VON STRUVE

Early in 1837 and just before Bessel began his intensive observations, Wilhelm von Struve from the Dorpat Observatory in Russia[2], announced what we might now term a "tentative" parallax for $\alpha$ Lyrae (Vega) of $0\overset{''}{.}125 \pm 0\overset{''}{.}055$ (von Struve, 1837). This $2.3\sigma$ detection closely matches the true parallax of $0\rlap{.}''13023 \pm 0\rlap{.}''00036$ (van Leeuwen, 2007)[3], although this close agreement is fortuitous given the large fractional uncertainty. Von Struve measured the position of Vega relative to a background star with a separation of 42". The parallax of that star (Gaia id = 2097892344993257344), not known at the time, has a very small value of 0.001463" +/- 0.000025" and thus did not significantly affect the (absolute) parallax measurement of Vega.

---

[2]Today, Dorpat is named Tartu and is part of Estonia

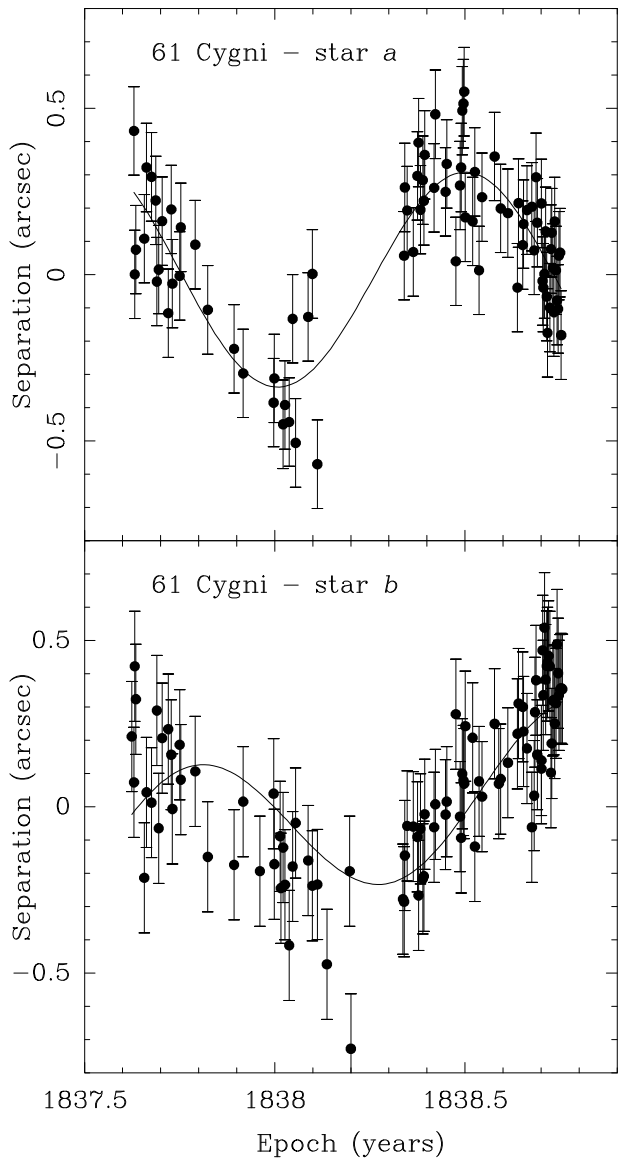

**FIGURE 1** Bessel's measurements of the angular separation of the midpoint of the 61 Cygni binary and two background stars (*a* and *b*). The solid lines are our fit allowing for parallax and (residual) proper motion, assuming uniform uncertainties for the measurements as Bessel did. While the data for star *a* fit the model well, the data for star *b* show clear systematic problems.

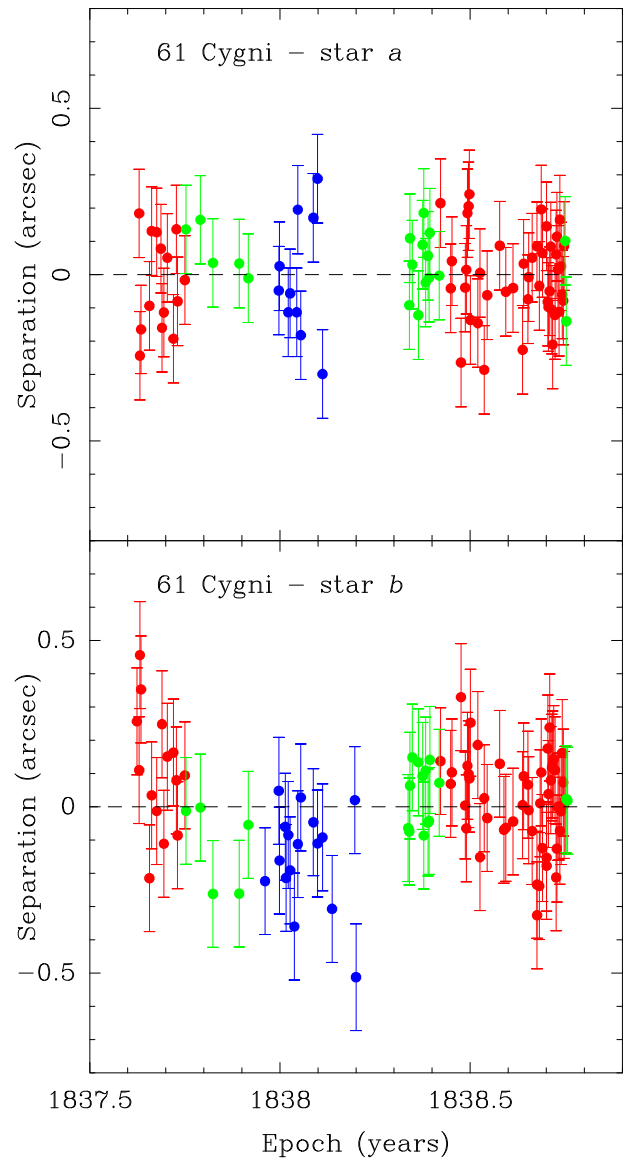

**FIGURE 2** Same as Fig. 1 but plotting residuals to the model fit. The points are color coded in *red* for warm-weather periods (0.42 to 0.75 decimal years), in *blue* for cold-weather periods (0.92 to 0.25), and in *green* for moderate-temperatures periods. Note that for star *b* the cold-weather points tend to have negative residuals and the warm-weather points tend to have positive residuals.

The raw data used by von Struve for his 1837 announcement is available in tabular form on page CLXXI in the voluminous book by von Struve (1837), whose $14^{\text{th}}$ chapter is devoted to his parallax measurement of Vega. Von Struve gives his model terms equated to his measured data in two coordinates: "distance" along the line between Vega and the background star and was published after Bessel's 1838 paper in the *Monthly Notices of the Royal Astronomical Society*. While von Struve's raw data starting from the beginning of 1837 are not included in his paper, his great-grandson Otto Struve, also a prominent astronomer, wrote articles in Sky and Telescope in 1956 about the first parallaxes. Struve (1956) plots the 1835–1838 separation distances and position angles on the sky between Vega and the background star versus time. We digitized these datasets and converted both to angular offsets in arcseconds. (Since this appears to be the only digitized record of von Struve's historic observations, we detail them in the Appendix.)

---

[3] Vega it is too bright for precision *Gaia* observations; this parallax is from the *Hipparcos* mission.

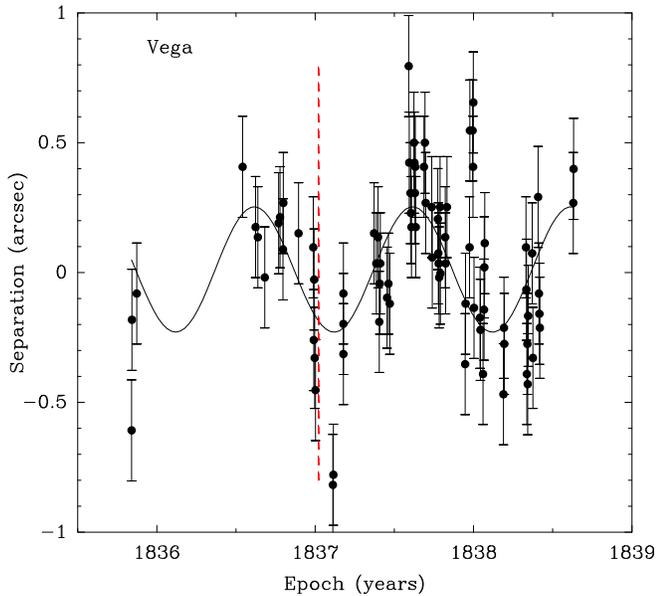

**FIGURE 3** Von Struve's parallax data for Vega. His tentative parallax of $0\rlap{.}{''}125$, announced in 1837, used data through the end of 1836 (indicated by the *red-dashed* line). He continued observations through late 1838 and revised his parallax to $0\rlap{.}{''}261$. The *solid* line shows our best-fitting parallax of $0\rlap{.}{''}271$, after removing an offset. The true parallax of Vega is $0\rlap{.}{''}130$.

Our re-analysis of the separation distances (Fig. 4) gives a parallax of $0\rlap{.}{''}271 \pm 0\rlap{.}{''}027$, very close to that quoted by von Struve. However, scaling the measurement uncertainties to achieve a post-fit $\chi_\nu^2$ of unity results in a larger parallax uncertainty of $\pm 0\rlap{.}{''}045$. Thus, it is likely that Struve adopted an optimistic estimate for the noise in his data. The direction data yield a smaller, but positive, parallax of $0\rlap{.}{''}077 \pm 0\rlap{.}{''}042$. Combining both distance and direction data yields a parallax of $0\rlap{.}{''}218 \pm 0\rlap{.}{''}031$ (see Table 2).

and "direction" perpendicular to that line.[4] On the same page, von Struve gives a "probable error" for his position measurements of $0\rlap{.}{''}155$. When we use this to fit his data, we find similar results (a parallax of $0\rlap{.}{''}129 \pm 0\rlap{.}{''}055$), but with a reduced chi-squared ($\chi_\nu^2$) of 2.17. In order to achieve a $\chi_\nu^2$ of unity, one needs to inflate his position uncertainties to $0\rlap{.}{''}228$, giving a more realistic parallax uncertainty of $\pm 0\rlap{.}{''}082$.[5]

It is interesting to fit von Struve's measurements in each coordinate separately. Using the "distance" measurements only, we find a parallax of $0\rlap{.}{''}252 \pm 0\rlap{.}{''}094$, whereas using the "direction" measurements yields $-0\rlap{.}{''}225 \pm 0\rlap{.}{''}165$. So, von Struve's tentative parallax in 1837 was essentially a weighted average of two results in significant tension ($2.5\sigma$), which balanced each other to give the "correct" result.

In October 1839, now with 96 measurements in hand, von Struve (1840)[6] claimed a parallax of Vega of $0\rlap{.}{''}261 \pm 0\rlap{.}{''}025$. This result is based on data taken between 1835 and 1838

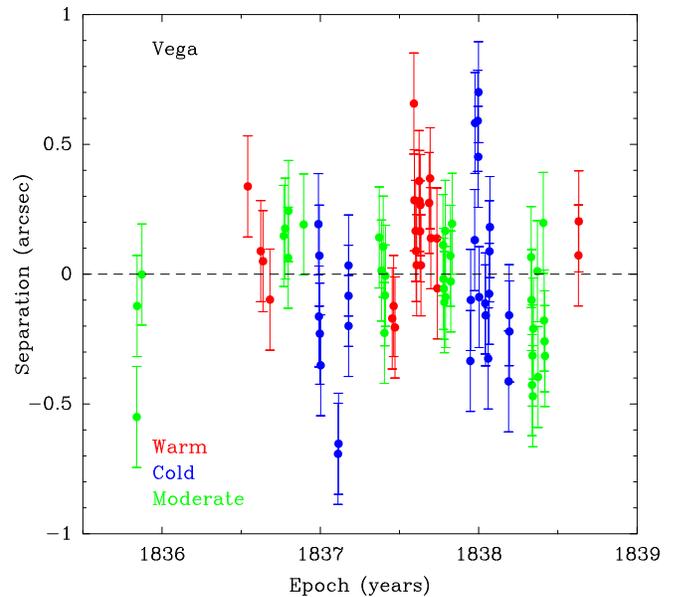

**FIGURE 4** Residuals for von Struve's "distance" data for Vega after removing the true parallax of $0\rlap{.}{''}130$ and the best fitting offset and motion. The points are color coded in 4-month bins in *red* for warm-weather periods (0.42 to 0.75 decimal years), in *blue* for cold-weather periods (0.92 to 0.25), and in *green* for moderate-temperatures periods. Note the tendency for the warm-weather points to have a positive bias and the cold-weather points a negative bias.

---

[4] We were able to reproduce von Struve's model parallax factors (i.e., expected offsets for a parallax of 1") by rotating expected parallax offsets calculated for the $(X, Y)$=(East,North) coordinates by $132°$ East of North, yielding $(X', Y')$, reversing the signs of $Y'$, and assigning $X'$ for the "distance" and $Y'$ for the "direction" model. The sign flip changes from our left- to his right-handed system. Note also that von Struve lists residuals from his best fit that use the (now) unconventional definition "model minus data".

[5] This is common practice today when one is uncertain of the magnitude of noise in a data set.

[6] In the NASA Astronomical Data System, various of F. G. W. von Struve's publications are erroneously credited to his son Otto Wilhelm von Struve.

**TABLE 2** Parallax of α Lyrae (Vega).

| Star | Data Used | Parallax von Struve | Parallax Re-analysis |
|---|---|---|---|
| Vega | 1835–1836 | $0.''125 \pm 0.''055$ | $0.''129 \pm 0.''082$ |
|  | Distances only | ... | $0.''252 \pm 0.''094$ |
|  | Directions only | ... | $-0.''225 \pm 0.''165$ |
| Vega | 1835–1838 | ... | $0.''218 \pm 0.''031$ |
|  | Distances only | $0.''261 \pm 0.''025$ | $0.''271 \pm 0.''045$ |
|  | Directions only | ... | $0.''077 \pm 0.''042$ |
|  | Temperature adjusted | ... | $0.''151 \pm 0.''058$ |

Comparison of parallax measurements of Vega by Friedrich Georg Wilhelm von Struve and our re-analyses of his data. The "Re-analysis" values allow for a residual proper motion and a constant measurement uncertainty, with the parallax uncertainty scaled to give a reduced chi-squared per degree of freedom of unity. The "Temperature adjusted" fit used both distances and directions and found temperature scaling parameters of $0.''082 \pm 0.''062$ for distance measurements and $0.''044 \pm 0.''059$ for direction measurements. The true parallax of Vega is $0.''130$.

This raises an interesting question: Why, with more observations, had von Struve discarded the "direction" data and only used "distance" measurements? In his paper, von Struve (1840) states the following:

"From my [96] measurements, the parallax could be determined in two different ways, from the observed separations or from the measured directions of the line connecting the two stars against the declination circle, the so-called position angles. However, since circumstances exist that impair the accuracy of the latter [measurements], these were not allowed to be used for the determination of the parallax and it was necessary that the parallax was derived from the separations alone."

Our re-fitting of the direction data leads to a parallax uncertainty that is comparable to that of the distance data, suggesting that the direction data was not of lesser precision. We can only speculate that the absense of a signficant parallax amplitude in the direction data led von Struve by 1840 to no longer trust that data.

Von Struve's 1840 result, even after inflating his uncertainty to achieve a $\chi_\nu^2$ of unity, departs significantly (by $3\sigma$) from Vega's true parallax of $0.''130$. Is there evidence in von Struve's data for a temperature effect, as we found for Bessel's data? In Fig. 4 we plot "distance" residuals after subtracting the effects of the true parallax of $0.''130$ and a best fitting constant and motion. These residuals show some systematic effects, with the warmest four months biased to positive values and the coldest four months biased to negative values. (We note that the cluster of points near 1838.3, which is within what normally is a cool-to-moderate temperature period at Dorpat Observatory's location, are also biased to negative values.) The average residual offset in warm months is $+0.''140$ while in cold months is $-0.''049$. This systematic difference correlates strongly with the peaks of the parallax sinusoid (see Fig. 3) and could lead to a falsely large fitted parallax by upwards of half their difference: $+0.''095$.

Knowing the true parallax of Vega, we could do a more rigorous investigation of systematics in von Struve's data owing to yearly temperature changes over the seasons. Lacking explicit temperature correction information from von Struve, we modeled the effects on measured distances and directions as yearly sinusoids, peaking in early July, and scaled by adjustable parameters to be determined in the least-squares fitting procedure. We held the parallax at its true value of $0.''130$ and found the temperature scaling parameter for distance data to be $0.''100 \pm 0.''034$. This value is in excellent agreement with our analysis in the previous paragraph, based on average biases seen in post-fit residuals. For the direction data we find a temperature scaling parameter of $0.''047 \pm 0.''059$, possibly indicating less sensitivity of the direction measurements to seasonal variations.

With stong evidence for significant temperature effects in von Struve's measurements, we performed a final least-squares fitting, allowing all parameters to vary simultaneously. This yields a parallax for the combined data of $0.''151 \pm 0.''058$ and distance and direction temperature parameters of $0.''082 \pm 0.''062$ and $0.''044 \pm 0.''059$, respectively. This parallax estimate is consistent with the true value for Vega and has $2.6\sigma$ formal statistical significance. The increased parallax uncertainty, compared to that with no temperature corrections, is owing to a large correlation coefficient of $-0.84$ between parallax and the distance data temperature parameter.

## 4 | THOMAS HENDERSON

Thomas Henderson observed $\alpha$ Centauri from a newly established observatory near Cape Town, South Africa. $\alpha$ Centauri is a triple star system which includes two bright stars, $\alpha^1$ Cen and $\alpha^2$ Cen, separated by $19''$. He reported transit measurements with a mural circle telescope of the *absolute* positions of both stars, calibrated by requiring that other stars observed on the same day had zero parallax. His paper, published in the *Memoirs of the Royal Astronomical Society*, gives tables of the right ascension and declination positions, as well as declinations from "reflected observations" (Henderson, 1840). Since the reflected observations had significantly larger uncertainties than the direct declinations, we do not consider them here.

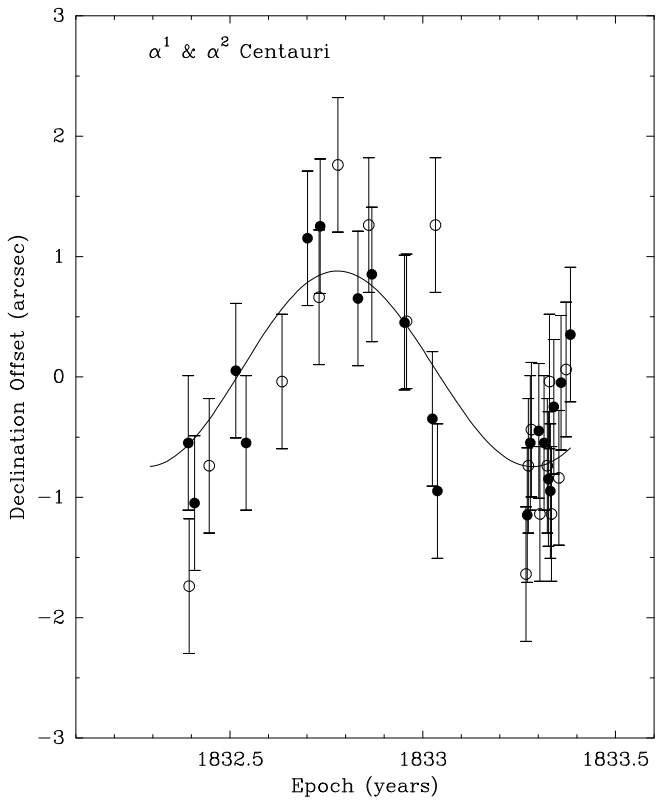

**FIGURE 5** Plot of Henderson's (more precise) declination data and the "Both Combined" parallax fit. A constant offset has been removed. Open circles are for $\alpha^1$ Cen and filled circles are for $\alpha^2$ Cen.

Parallaxes measured from absolute positions (i.e., not relative to nearby stars measured concurrently as did Bessel and von Struve) are directly affected by corrections for aberration of light, which also is owing to the orbital motion of the Earth and has a yearly period. Henderson notes that his tabular data assumed a constant of aberration of $20.''50$, virtually identical to the modern value[7]. He also comments on the probable uncertainties for measurements of right ascension ($\pm 0.''71$) and declination ($\pm 0.''38$). However, for the more accurate, and hence more important, declination data he notes that experience from his previous "Catalog of declinations" (Henderson, 1837) suggested a larger declination uncertainty of $\pm 0.''52$ was appropriate for the typical zenith distance of 26° of $\alpha$ Cen, and he adopts the average of these two declination uncertainties: $\pm 0.''45$.

Henderson's data and our parallax fit are plotted in Fig. 5 and listed in Table 3. In most cases, there are reasonably good correspondences between Henderson's parallax values and our re-fitted ones. One exception is for the right ascension parallax for $\alpha^2$ Cen. In some cases, Henderson's and our parallax uncertainties agree within 10%. However, there are several examples where our uncertainties are upwards of 50% larger than Henderson's. These differences can be accounted for by scaling of measurement uncertainties to achieve a reduced $\chi^2_\nu$ of unity. Thus, it appears that Henderson's estimates of measurement uncertainties are likely optimistic in some cases.

All in all, it is clear that Henderson's final parallax estimate, combining the right ascension and declination results for both stars was precise and formally significant at the $7\sigma$ level. However, the true parallax of $\alpha$ Cen, as measured by the *Hipparcos* mission (van Leeuwen, 2007), is $0.''755 \pm 0.''004$.[8] Thus, our re-analysis of Henderson's data, which gives a parallax of $1.''09 \pm 0.''15$, is $2.2\sigma$ from the true value. This moderate tension with the true parallax suggests some small residual systematics in his data.

## 5 | CONCLUSIONS

We have performed an in-depth re-analysis of the datasets that Bessel, von Struve, and Henderson used to obtain the first stellar parallaxes, and we can reproduce their results. It is clear that von Struve and Henderson used *a priori* estimates of data errors and arrived at optimistic parallax uncertainties. We have re-scaled their data errors to achieve post-fit $\chi^2_\nu$ per degree of freedom of unity and arrived at more realistic parallax uncertainties. Even based on more realistic uncertainties, we can confirm that all three astronomers detected their stellar parallaxes with reasonable statistical significance.

We confirm Bessel's 1838 parallax estimates for 61 Cygni in detail. Bessel measured parallaxes relative to two reference

---

[7] Henderson also presents parallax results adopting a value of $20.''36$ for the constant of aberration. Using this value changes his parallax results for $\alpha^1$ Cen and $\alpha^2$ Cen by 12% and 26% for right ascension and by 1% and 3% for declination data.

[8] These stars are too bright for precision *Gaia* observations.

**TABLE 3** Parallax of $\alpha$ Centauri.

| Star | Data Used | Parallax Henderson | Parallax Re-analysis |
|---|---|---|---|
| $\alpha^1$ Cen | Right Ascension | $0''.82 \pm 0''.35$ | $0''.80 \pm 0''.49$ |
| ... | Declination | $1''.40 \pm 0''.19$ | $1''.37 \pm 0''.27$ |
| ... | Combined | $1''.27 \pm 0''.17$ | $1''.25 \pm 0''.23$ |
| $\alpha^2$ Cen | Right Ascension | $0''.38 \pm 0''.34$ | $0''.47 \pm 0''.51$ |
| ... | Declination | $1''.02 \pm 0''.18$ | $1''.00 \pm 0''.19$ |
| ... | Combined | $0''.88 \pm 0''.16$ | $0''.94 \pm 0''.18$ |
| Both | Combined | $1''.06 \pm 0''.12$ | $1''.09 \pm 0''.15$ |

Comparison of parallax measurements of $\alpha$ Cen by Thomas Henderson and our re-analysis of his data. Henderson's values are given for the constant of aberration of $20''.50$ and assume that proper motion had been removed from the data. For comparison, the modern parallax of $\alpha$ Cen is approximately $0''.75$.

stars (*a* and *b*). These two parallax estimates were in some tension and Bessel, noting the difference, suggested the measurement with the smaller parallax (star *b*) could be explained if that reference star itself had a significant parallax. However, the *Gaia* space mission has measured the parallax of both reference stars and found them to be quite small, discounting Bessel's suggestion. Upon closer inspection of the data, we find that star *b* has residuals that display systematic trends, which correlate well with his recorded temperature corrections. Doubling his temperature corrections removes the problems seen in the systematic residuals. In 1840 with more data, Bessel published improved parallaxes, which included prescriptions for the effects of temperature corrections on parallax estimates. With perfect hindsight, knowing the true parallax of 61 Cygni to be $0''.286$, we verify the need for large temperature corrections, which then yield excellent agreement between the parallaxes measured against each reference star.

It has long been a mystery as to why von Struve claimed a parallax for Vega of $0''.125$, which is nearly identical to its true value, only later with more measurements to arrive at a parallax of $0''.261$. His early result comes from combining two dimensions of measurements: along and perpendicular to the position angle of the reference star. These two measurement sets yield substantially different parallaxes, but the differences from the true parallax fortuitously cancel when a weighted average is performed. Von Struve later discarded the perpendicular data, only using the "more precise" distances between Vega and the reference star for a parallax fit. As we found for Bessel, this dataset also shows some residual systematics that are likely from temperature-dependent effects. Accounting for these effects yields a parallax value of $0''.151 \pm 0''.058$, which is statistically consistent with the true parallax of $0''.130$.

We note that Bessel and von Struve also made significant contributions to the study of wide binaries. Since these measurements are still used to provide a long time-baseline to complement current observations, our finding of temperature sensitivities in their astrometric data might be of importance for studies of long period binaries.

In general, we can reproduce Henderson's analyses for $\alpha$ Centauri, although in some cases scaling the measurement errors to achieve a $\chi^2_\nu$ of unity results in larger parallax uncertainties. Accounting for this, we find a best-fit parallax of $1''.09 \pm 0''.15$, which is only in mild tension with the true value of $0''.75$.

Remarkably, after a century-long quest for the first stellar parallax, within the brief period of just 3 years, three astronomers met success and obtained results that have stood the test of time.

## ACKNOWLEDGMENTS

We would like to thank Arne Hoyer, who traced a paper copy of F. G. W. von Struve's monograph *Mensurae micrometricae* to the Bavarian State Library and to that fine institution for providing a digital copy of this work's chapter 14. We are thankful to Jens Kauffmann for his expert help with LaTeX formatting issues.

# 6 | APPENDIX

Table 4 lists Wilhelm von Struve's measurements of the separation ("distance") between Vega and a reference star and the perpendicular offset ("direction"). The data were hand digitized from plots in a paper by Otto Struve in Sky & Telescope (1956). We subtracted 5.616 from the "distance" data and scaled them by 7.75 to convert to arcsecond offsets. For the "direction" data, originally listed as position angles, we subtracted $138°\!.25$, converted to radians, and multiplied by the separation between stars of $42''\!.016$.

**TABLE 4** Wilhelm Struve's data for Vega.

| Date | Distance (″) | Direction (″) | Date | Distance (″) | Direction (″) |
|---|---|---|---|---|---|
| 1835.841 | −0.612 | −0.308 | 1837.736 | 0.248 | 0.345 |
| 1835.843 | −0.186 | −0.353 | 1837.739 | 0.054 | 0.038 |
| 1835.873 | −0.085 | −0.278 | 1837.775 | 0.202 | 0.270 |
| 1836.542 | 0.403 | −0.443 | 1837.778 | 0.070 | 0.315 |
| 1836.624 | 0.171 | −0.210 | 1837.780 | 0.031 | 0.225 |
| 1836.638 | 0.132 | −0.263 | 1837.783 | −0.023 | 0.135 |
| 1836.682 | −0.023 | 0.068 | 1837.789 | 0.248 | ... |
| 1836.769 | 0.186 | −0.038 | 1837.791 | −0.008 | 0.060 |
| 1836.778 | 0.209 | −0.293 | 1837.821 | 0.132 | 0.135 |
| 1836.797 | 0.085 | −0.248 | 1837.824 | 0.031 | 0.188 |
| 1836.799 | 0.264 | 0.015 | 1837.832 | 0.248 | 0.195 |
| 1836.895 | 0.147 | 0.113 | 1837.947 | −0.357 | 0.128 |
| 1836.988 | 0.093 | −0.315 | 1837.950 | −0.124 | 0.360 |
| 1836.991 | −0.264 | −0.173 | 1837.975 | 0.093 | −0.120 |
| 1836.994 | −0.031 | −0.210 | 1837.977 | 0.543 | −0.150 |
| 1836.997 | −0.333 | −0.563 | 1837.994 | 0.543 | −0.210 |
| 1837.002 | −0.457 | 0.083 | 1837.997 | 0.403 | −0.045 |
| 1837.112 | −0.822 | −0.338 | 1837.999 | 0.651 | 0.098 |
| 1837.115 | −0.783 | −0.128 | 1838.003 | −0.140 | 0.285 |
| 1837.176 | −0.202 | −0.173 | 1838.041 | −0.178 | 0.293 |
| 1837.178 | −0.318 | 0.315 | 1838.044 | −0.225 | 0.345 |
| 1837.180 | −0.085 | −0.150 | 1838.060 | −0.395 | 0.233 |
| 1837.372 | 0.147 | −0.218 | 1838.066 | −0.147 | 0.443 |
| 1837.378 | ... | 0.278 | 1838.068 | 0.016 | 0.413 |
| 1837.383 | ... | −0.113 | 1838.071 | 0.109 | 0.263 |
| 1837.386 | 0.031 | 0.315 | 1838.189 | −0.473 | 0.053 |
| 1837.392 | ... | 0.098 | 1838.192 | −0.217 | 0.225 |
| 1837.397 | 0.132 | 0.353 | 1838.194 | −0.279 | 0.360 |
| 1837.405 | −0.194 | 0.653 | 1838.197 | ... | 0.638 |
| 1837.408 | −0.047 | 0.413 | 1838.326 | ... | 0.188 |
| 1837.411 | 0.031 | −0.060 | 1838.329 | ... | 0.661 |
| 1837.454 | −0.101 | 0.015 | 1838.331 | 0.093 | 0.353 |
| 1837.463 | −0.047 | −0.038 | 1838.334 | −0.070 | 0.713 |
| 1837.471 | −0.124 | 0.083 | 1838.337 | −0.395 | 0.488 |
| 1837.591 | 0.791 | 0.285 | 1838.340 | −0.279 | 0.518 |
| 1837.594 | 0.419 | 0.383 | 1838.342 | −0.434 | 0.593 |
| 1837.602 | 0.302 | 0.188 | 1838.345 | −0.171 | 0.736 |
| 1837.605 | 0.225 | 0.435 | 1838.370 | 0.070 | 0.398 |
| 1837.608 | 0.171 | 0.240 | 1838.375 | −0.333 | 0.668 |
| 1837.624 | 0.496 | 0.038 | 1838.405 | ... | 0.248 |
| 1837.627 | 0.419 | 0.180 | 1838.408 | 0.287 | 0.300 |
| 1837.630 | 0.302 | 0.075 | 1838.413 | −0.085 | 0.420 |
| 1837.632 | 0.403 | 0.390 | 1838.416 | −0.163 | 0.608 |
| 1837.635 | 0.171 | 0.173 | 1838.419 | −0.217 | 0.330 |
| 1837.687 | 0.403 | 0.075 | 1838.422 | ... | 0.188 |
| 1837.693 | 0.496 | 0.135 | 1838.630 | 0.264 | 0.315 |
| 1837.698 | 0.264 | 0.165 | 1838.632 | 0.395 | 0.383 |

Columns are (twice): year, "distance" and "direction" data for $\alpha$ Lyr (see text); taken from Otto Struve (Sky & Telescope, 1956).